\documentstyle[11pt,epsf]{article}

\begin{document}
\title{Rotating wave approximation: systematic expansion and 
application to coupled spin pairs}

\author{Bernhard Thimmel, Peter Nalbach\\
Institut f\"ur Theoretische Physik, Universit\"at Heidelberg\\
Philosophenweg 19, D-69120 Heidelberg, Germany\\
\\
Orestis Terzidis\\
Service de Physique de l'Etat Condens\'e, C.E. Saclay\\
Orme des Merisiers, 91191 Gif-sur-Yvette Cedex, France}
\date{\today}
\maketitle

\begin{abstract}
We propose a new treatment of the dynamics of a
periodically time-dependent Liouvillian by
mapping it onto a time-independent problem and
give a systematic expansion for its effective Liouvillian.
In the case of a two-level system, the lowest order contribution 
is equivalent to the well-known rotating wave approximation. We 
extend the formalism to
a pair of coupled two-level systems. For this pair, we find
two Rabi frequencies and we can give parameter regimes
where the leading order of the expansion is suppressed and
higher orders become important. These results might help
to investigate the interaction of tunneling systems in mixed crystals
by providing a tool for the analysis of echo experiments.
\end{abstract}
\bigskip
PACS numbers: 76.60.L , 66.35.+a
\section{Introduction}

Quantum tunneling of substitutional defect ions in
alkali halide crystals leads to particular low
temperature properties \cite{Nara,Aloisbuch}. Due to their misfit
in size or shape such defect ions are confined to a 
potential energy landscape with a few degenerate
potential wells. At low temperatures thermally
activated hopping is inhibited and
the defect ion passes through the barrier by quantum
tunneling. These tunneling defects are usually modeled by 
two-level-systems (TLSs), which dynamically are identical to spin 
degrees of freedom. 

Depending on the
detailed nature of the TLS, both electromagnetic and acoustic 
fields might couple to the system.
One of the interesting features of such systems is their 
response to an oscillating weak external field which is in resonance
with the TLS.
The key point in this context is the emergence 
of the Rabi oscillation \cite{sli}, which is related to the 
amplitude of the external field and which allows a systematic 
experimental investigation of the response of a TLS.

The easiest way to derive the Rabi oscillation is to do the so called
{\it rotating wave approximation} (RWA) as originally introduced by 
Rabi \cite{RabOsc}. Since then, the problem has been tackled 
by numerous methods starting with the work of Bloch and Siegert
\cite{BSshift}, who could first give a correction to the RWA.
Further approaches have been a solution in terms of continued
fractions presented by Autler and Townes \cite{Contfrac} and
especially the Floquet formalism developed by Shirley \cite{Floquet}. 

In this paper, we present yet another expansion scheme for solving
the problem. It is somewhat related to the Floquet formalism, but we try
to find an effective static Liouvillian or Hamiltonian rather than 
a time-evolution operator, which makes our calculations easier.
Like the Floquet formalism, our approach can be easily generalised
to systems involving several energy levels, where the simple
geometrical interpretation of the RWA fails. Thus, we can define
the RWA in these situations as the lowest order contribution of
our expansion scheme.

Our paper is organised as follows. In the following section, we
address the time-dependent problem of a single spin. We first
rederive the RWA, then in section \ref{Formalismus}, we present the
formalism for our expansion. In the third section then, we use our
new method for treating the problem of two coupled spins. The main
results of this section are the emergence of two Rabi frequencies,
whose relative magnitudes depend on the spin-spin coupling, and
the identification of parameter regimes where higher order corrections
to the RWA become important.

\section{The treatment of a single spin}
This section deals with the problem of one TLS that is coupled
to an external force. The TLS is modeled by the Hamiltonian 
\cite{Nara,Aloisbuch}
\begin{equation} 
	H_{\rm sys} = -k \frac{\sigma_x}{2} - \Delta \frac{\sigma_z}{2}\ ,
\end{equation}
with tunneling element $k$, asymmetry $\Delta$, and energy
$ \epsilon = \sqrt{k^2 + \Delta^2}$.
The external force is given by
\begin{equation}
	H_{\rm ext} =	- 2 \eta \, \sin(\omega t) \frac{\sigma_z}{2}\ .
\end{equation}
This choice of $H_{\rm ext}$ implies that the external field couples 
solely to the spatial coordinate $\sigma_z$ of the TLS, 
as realized in the case of an electric field coupling to 
the dipole moment of the TLS.

The time evolution of the statistical operator
is governed by the Bloch equation \cite{Blochgl,sli} 
\begin{equation} \label{IntBloch}
	\frac{d\rho}{dt} = i [\rho,H_{\rm sys} + H_{\rm ext}] 
		- \gamma( \rho -\rho^{eq}) =: {{\cal L}} \rho\ ,
\end{equation} 
where the right hand side of the equation defines the Liouvillian 
${\cal L}$. The second term in the Bloch equation describes a 
relaxation of the statistical operator towards its
equilibrium given by 
$\rho^{eq} = e^{-\beta H_{\rm sys}} / \left(\mbox{tr}\; e^{-\beta 
H_{\rm sys}} \right)$. This choice of $\rho^{eq}$ is justified
as long as the relaxation time is much longer than the period 
of the driving field, $\gamma \ll \omega$.
If we expand the density matrix
in terms of the Pauli matrices and the $\bf 1$-matrix,
\begin{equation}
	\rho = \frac 1 2 ( {\bf 1} + r_x \sigma_x + r_y \sigma_y + r_z 
			\sigma_z)
	=: \left(\begin{array}{c}
			1  \\
			r_x\\
			r_y\\
			r_z\\
		\end{array}\right)\ , \qquad r_x^2 + r_y^2 + r_z^2 \le 1\ ,
\end{equation}
the Liouvillian ${\cal L}$ can be represented by a $4\times4$ matrix. 
Eq. (\ref{IntBloch}) then yields
a differential equation for the vector $(1,r_x,r_y,r_z)$.
\subsection{RWA}
Let us first rederive the RWA.
To simplify the problem, we neglect the asymmetry and consider
\begin{equation} \label{SingHam}  
	H = -\epsilon \frac{\sigma_x}{2} - 2 \eta \, \sin(\omega t) 
	\frac{\sigma_z}{2}\ .
\end{equation}
The Bloch equation in this case reads 
\begin{equation} \label{SingBloch}
	\frac{d\rho}{dt} = i [\rho,H] - \gamma( \rho -\rho^{eq})\ , 
\end{equation} 
and $\rho^{eq}$ has the particularly simple form
\begin{equation}
	\rho^{eq} = \frac 1 2 ( 1 + n^0 \sigma_x),
	\qquad n^0=\tanh(\beta \epsilon/2)\ .
\end{equation}
Throughout this section, we assume that both the coupling to the external
field and the damping are small and that we are close to resonance, i.e.
\begin{equation}\label{SingAss}
	\eta, \gamma, |\epsilon-\omega| \ll \omega \ .
\end{equation}

The Hamiltonian in (\ref{SingHam}) describes not only a 
tunneling system coupled to a harmonic oscillator, but equally 
well a spin--$\frac 1 2$ particle subjected to a static magnetic field in 
$x$--direction, $B_x = \epsilon/(g\mu_B)$, and an oscillating field in 
$z$--direction, $B_z = 2\eta/(g\mu_B)\sin(\omega t)$. In the latter
interpretation, the components of the density matrix $(r_x,r_y,r_z)$
are proportional to the magnetic moment of the spin. Thus, under rotation
they transform like a pseudo-vector. In the following we will use the 
terminologies of both interpretations.

As a preliminary, we consider the case of a static magnetic field.
From the classical analogue one can infer that the magnetic moment
will precess round the direction of the magnetic field with a 
frequency proportional to the field strength. 
Thus, the static Hamiltonian
\begin{equation}
	H = \epsilon \left( e_x \frac {\sigma_x}{2} + e_y \frac{\sigma_y}{2} 
		+ e_z \frac{\sigma_z}{2}\right)\ ,\qquad 
		e_x^2 + e_y^2 + e_z^2 = 1
\end{equation}
describes a rotation of the vector $(r_1,r_2,r_3)$ about the axis 
$(e_x,e_y,e_z)$ with an angular frequency $\epsilon$.
The explicit formula for the statistical operator as a function of
time is useful in the following, we give it for $\gamma=0$, i.e.
for the case without dissipation:
$\rho(t) = {\cal U}_0(t) \rho(0) = e^{-iHt} \rho(0) e^{iHt}\ ,$
where the time evolution operator ${\cal U}_0(t)$ can be represented 
by the matrix
\begin{equation} \label{SingCumb} {\cal U}_0(t) = 
\left( \begin{array}{cccc} 1 & 0 & 0 & 0\\
	0 & c\! +\! e_x^2 (1\!-\!c) & e_x e_y (1\!-\!c)\! -\! e_z s &
	e_x e_z (1\!-\!c)\! + \!e_y s \\
	0 & e_x e_y (1\!-\!c)\! +\! e_z s & c\! +\! e_y^2 (1\!-\!c) &
	e_y e_z (1\!-\!c)\! -\! e_x s \\
	0 & e_x e_z (1\!-\!c)\! -\! e_y s &
	e_y e_z (1\!-\!c)\! + \!e_x s & c\! +\! e_z^2 (1\!-\!c) \\	
\end{array}\right)\ ,	
\end{equation}
using the abbreviations $c := \cos(\epsilon\,t),\  s := \sin(\epsilon\,t)$.

In the presence of dissipation, the corresponding solution reads
\begin{equation}\label{SingSst}
	\rho(t) = {\cal U}(t)\rho(0) = \rho^{\rm st} + e^{-\gamma t} \, 
	{\cal U}_0(t) \!
		\left(\rho(0) - \rho^{\rm st}\right)\ ,
\end{equation}
where the stationary solution $\rho^{\rm st}$ is given by
${\cal L}\, \rho^{\rm st} = 0$.

\label{RWA}
Now consider the Hamiltonian (\ref{SingHam}), which includes 
non-static terms. We split it into two parts
\begin{eqnarray}\label{SingAufspaltung}
	&& H\  = \  H_0 + H_1 \nonumber\\
	&&\ H_0\  := \ -\omega \, \frac{\sigma_x}{2} \nonumber\\
	&&\ H_1\  := \ -\delta \, \frac{\sigma_x}{2} - 2 \eta \, 
		\sin(\omega t) \,
		\frac{\sigma_z}{2} \qquad,\qquad \delta := \epsilon-\omega\ ,
\end{eqnarray}
where assumption (\ref{SingAss}) ensures that $H_1$ 
is a weak perturbation to $H_0$.

The dominant part $H_0$ of the Hamiltonian leads to a rotation of the
magnetic moment about the $x$-axis. We take this into account by 
changing to a rotating frame of reference.
In this rotating reference frame, we denote the statistical operator by
$\bar{\rho}$. It is related to $\rho$ through
\begin{equation}
	\rho(t) = e^{-iH_0t} \bar{\rho}(t) e^{iH_0t} = 
		e^{i\omega t (\sigma_x/2)} \bar{\rho}(t) 
		e^{- i\omega t (\sigma_x/2)}
\end{equation}
and its equation of motion reads
\begin{eqnarray}\label{SingeffBloch}
	\frac{d\bar{\rho}}{dt} 	& = & 
		i [\bar{\rho},\bar{H}] - \gamma(\bar{\rho}-\rho^{eq})
		=: \bar{{\cal L}} \bar{\rho}\ ,
\end{eqnarray}
with the transformed Hamiltonian
\begin{equation}\label{SingtrHam}
   \bar{H} = - \delta \, \frac{\sigma_x}{2} + \eta \,\frac{\sigma_y}{2} 
		- \eta \, 
		e^{-i2\omega t (\sigma_x/2)} \, \frac{\sigma_y}{2} \, 
		e^{i2\omega t (\sigma_x/2)}\ .
\end{equation}
The form of the 
effective Hamiltonian is easily understood by observing that the linearly 
polarised field $H_{\rm ext}$ can be decomposed into two counterrotating 
circularly polarised fields of strength $(B_z/2)$ and angular frequency
$\omega$ each. In the rotating frame of reference, one of them becomes a 
static field whereas the frequency of the other one is doubled.

In the following we want to neglect the oscillating field. This
is justified by the following consideration.
During one period of the oscillating field, $\bar{\rho}$ changes only 
little, as its typical time constant is 
$1/(\sqrt{\delta^2 + \eta^2}) \gg (1/2\omega)$. Therefore,
it is plausible to assume that the contributions of the 
time dependent field cancel over a period.
This approximation is called {\it rotating wave approximation} because
the original sinusoidal magnetic field is replaced by
a rotating field of half the amplitude and the same frequency.
The remaining part of the Hamiltonian is static, 
\begin{equation}\label{SingeffHam}
   \bar{H}_{\rm eff} = - \delta \, \frac{\sigma_x}{2} + \eta \,
		\frac{\sigma_y}{2} \ ,
\end{equation}
describing a rotation with angular velocity $(-\delta,\eta,0)$.
Thus, we can solve the equations of motion (\ref{SingeffBloch}) using
(\ref{SingSst}) while the stationary solution reads
\begin{equation}
	\rho^{\rm st} = \frac{1}{2} \left[{\bf 1} + \frac{n^0}{\delta^2 
		+ \eta^2 + \gamma^2}
		\left\{ (\delta^2 + \gamma^2)\, \sigma_x - \eta\delta \, \sigma_y 
		-\eta \gamma \, \sigma_z \right\} \right], \quad 
		n^0 = \tanh (\beta \epsilon/2)\ .
\end{equation}
If at $t=0$, the system is in equilibrium, we find
for the dipole moment of the TLS in the laboratory frame
\begin{eqnarray}
	\left\langle\hat{p}\right\rangle (t) = \mbox{tr}\; ( \hat{p} \rho(t))
	= - n^0 \frac{\eta}{\tilde{\Omega}}&&\hspace{-6mm}\left[\,
	\cos (\omega t) \left( \frac{\gamma}{\tilde{\Omega}} + e^{-\gamma t}
		\left\{ \frac{\Omega}{\tilde{\Omega}} \sin(\Omega t) - 
			\frac{\gamma}{\tilde{\Omega}} \cos(\Omega t)\right\}\right)
		\right. \nonumber\\
&&	\hspace{-4mm}\left. +\sin(\omega t)\left( -\frac{\delta}{\tilde{\Omega}} 
	+ e^{-\gamma t}
		\left\{ \frac{\delta}{\tilde{\Omega}} \cos(\Omega t) + 
			\frac{\delta\gamma}{\Omega\tilde{\Omega}} \sin(\Omega t)\right\}
	\right)\right]
\end{eqnarray}
using the definitions
$\tilde{\Omega}^2:=\Omega^2 + \gamma^2 := \delta^2 + \eta^2 + \gamma^2$;
$\Omega$ is called {\it Rabi frequency}. Fourier transformation 
yields peaks at $\omega$ and $\omega\pm\Omega$ whose heights have 
an overall factor of $(\eta/\tilde{\Omega})$, ensuring that we 
only find a signal close to resonance.
 
Although the method of the rotating wave approximation is intuitively very
appealing, it lacks a rigorous foundation. In particular, there are no
means to estimate the quality of the approximation and its
limits of validity.

\subsection{An expansion scheme for the RWA}\label{Formalismus}
In this section, we present a systematic expansion, which yields 
the rotating wave approximation as the lowest order contribution.
We consider the Hamiltonian
\begin{equation} \label{FormHama}  
	H = -k \frac{\sigma_x}{2} - \Delta \frac{\sigma_z}{2} - 2 \eta 
		\sin(\omega t) \frac{\sigma_z}{2} \ ,
\end{equation}
where only the strength of the driving field $\eta$ and the damping 
constant of the Bloch equation $\gamma$ 
are small compared to $\omega$:
\begin{equation}\label{FormAss}
	\eta, \gamma \ll \omega.
\end{equation}
The time evolution is dominated by the static part of the Hamiltonian.
Therefore we start by diagonalizing this part by means of a rotation
about the $y$-axis,
\begin{eqnarray} \label{FormHam}  
	\hat{H} & = & e^{- i \phi (\sigma_y/2)} \, H \, e^{i 
		\phi(\sigma_y/2)}\nonumber\\
	& = & -\epsilon \frac{\sigma_x}{2} - 2 u \eta \, \sin(\omega t) 
		\frac{\sigma_z}{2} 
		- 2 v \eta \, \sin(\omega t) \frac{\sigma_x}{2} \ ,
\end{eqnarray}
where
\[
	\epsilon=\sqrt{k^2+\Delta^2},\qquad \phi=\arctan(k/\Delta), \qquad
	u=\cos(\phi),\qquad v=\sin(\phi)\ .
\]
After this unitary transformation, the equilibrium value of the 
statistical operator reads
\begin{equation} 
	\rho^{eq} = \frac 1 2 \left[ {\bf 1} + n^{0} \sigma_x \right], 
		\quad n^0 = \tanh(\beta\epsilon/2)\ .
\end{equation}
 
Now, we proceed in a similar way as in the last section. The 
Hamiltonian is divided into two parts
\begin{eqnarray}\label{FormAufsp}
	&& \hat{H}\  = \  H_0 + H_1 \nonumber\\
	&&\ H_0\  := \ -n \omega \, \frac{\sigma_x}{2} \nonumber\\
	&&\ H_1\  := \ -\delta \, \frac{\sigma_x}{2} - 2 u \eta \, 
		\sin(\omega t) \frac{\sigma_z}{2}
		- 2 v \eta \, \sin(\omega t) \frac{\sigma_x}{2}
		,\qquad \delta := \epsilon-n \omega\ ,
\end{eqnarray}
where the integer $n$ has to be chosen such that 
$|\delta| < \frac{\omega}{2}$ thereby ensuring that $H_1$ is small
compared to $\omega$.
The effect of $H_0$ can be taken into 
account by changing to a frame of reference spinning around the $x$-axis
with the frequency $|n\omega|$
\begin{eqnarray}
	\rho(t) & = & e^{-iH_0t} \bar{\rho}(t) e^{iH_0t} = 
		e^{i n\omega t (\sigma_x/2)} \bar{\rho}(t) 
		e^{- i n\omega t (\sigma_x/2)}\nonumber\\
	\frac{d\bar{\rho}}{dt} & = & 
		i [\bar{\rho},\bar{H}] - \gamma(\bar{\rho}-\rho^{eq})
		=: \bar{{\cal L}} \bar{\rho} \ ,
\end{eqnarray}
where $\bar{{\cal L}}$ reads in the
basis $(1,\sigma_x,\sigma_y,\sigma_z)$
\begin{equation}
	\bar{{\cal L}}(t)  =  \left( \begin{array}{cccc}
		0 & 0 & 0 & 0 \\
		\gamma n^0 & -\gamma & 
			-u \eta \,g(t) &
			u \eta \,f(t)\\
		0 & u \eta \,g(t) &
		-\gamma & \delta + 2 v \eta \, \sin(\omega t) \\
		0 & -u \eta \,f(t) &
		- ( \delta + 2 v \eta \, \sin(\omega t)) & -\gamma
		\end{array}\right)\ ,
\end{equation}
using the definitions
\begin{eqnarray}
	f(t) & = & \cos((n-1)\,\omega t) - \cos((n+1) \,\omega t)\nonumber\\
	g(t) & = & \sin((n-1)\,\omega t) - \sin((n+1) \,\omega t)\nonumber\ .
\end{eqnarray}
In the following, we
first show that there is an effective, static Liouville operator 
yielding a time evolution that matches the correct time evolution 
at integer multiples of the period $\tau:=2\pi/\omega$.
Then, we give an expansion
of the effective Liouville operator in powers of $(1/\omega)$.

If we use the time ordering operator $\cal T$, the formal solution 
for the evolution of $\bar{\rho}$ reads
\begin{equation}
	\bar{\rho}(t) = {\cal T} \exp\left[\int_{0}^{t}\bar{{\cal L}}(t')
		\;dt'\right]
		\bar{\rho}(0)\ .
\end{equation}
Taking advantage of the periodicity of $\bar{{\cal L}}(t)$, we 
rewrite the last equation as
\[
	\bar{\rho}(t) = \left\{ {\cal T} \exp\left[\int_{N\tau}^{t}
		\bar{{\cal L}}(t')\;dt'
		\right]\right\} \left\{ {\cal T} \exp\left[\int_{0}^{\tau}
		\bar{{\cal L}}(t')\;dt'
		\right]\right\}^N\bar{\rho}(0),\quad N \tau < t < (N+1) \tau \ .
\]
Now, we define the effective Liouvillian ${\cal L}_{\rm eff}$ by 
\begin{equation} 
	\exp\left[{\cal L}_{\rm eff} \tau\right] = 
		{\cal T} \exp\left[\int_{0}^{\tau}\bar{{\cal L}}(t')\;dt' \right]
\end{equation}
and end up with
\begin{equation}
	\bar{\rho}(t) = \left\{ {\cal T} \exp\left[\int_{N\tau}^{t}
		\bar{{\cal L}}(t')\;dt'
		\right]\right\} \exp\left[{\cal L}_{\rm eff} N \tau
		\right]\bar{\rho}(0)\ .
\end{equation}
Thus, we separated the problem into two parts: first, one has to 
calculate the evolution between $N\tau$ and $t$ accessible by simple 
perturbation theory. Second, one has to determine ${\cal L}_{\rm eff}$. 
We discuss the latter first.

We use the identity
\begin{eqnarray}\label{FormId}
	\exp[{\cal L}_{\rm eff} N \tau]
	& = & {\cal T} \exp\left[\int_{0}^{N\tau}\bar{{\cal L}}(t')
		\;dt'\right]\nonumber\\
	\hspace{-5mm}\Longleftrightarrow\quad 
	{\cal L}_{\rm eff} N \tau + \frac 1 2 {\cal L}_{\rm eff}^2 
		(N \tau)^2 + \ldots
	& = & \int\limits_{0}^{N\tau} \! dt_1\,\bar{{\cal L}}(t_1) + 
		\int\limits_{0}^{N\tau} \! dt_1\!\!
		\int\limits_{0}^{t_1} \!dt_2 \, \bar{{\cal L}}(t_1) 
		\bar{{\cal L}}(t_2) + \ldots\ ,
\end{eqnarray}
which implies that ${\cal L}_{\rm eff}$ is the sum of all terms 
on the right hand side 
proportional to $N\tau$. Luckily, these terms form a series in 
$(1/\omega)$ that can be calculated explicitly order by order.
By substituting $z_i := \omega t_i$, the $m^{\rm th}$ integral yields
\begin{equation} 
	\frac{1}{\omega^m} \int\limits_{0}^{2 \pi N} \! dz_1\!\!\int
		\limits_{0}^{z_1} 
		\!dz_2\ldots\!\!\!\!\int\limits_{0}^{z_{m-1}} \!dz_m \, 
		\bar{{\cal L}}(z_1/\omega) \bar{{\cal L}}(z_2/\omega) 
		\ldots \bar{{\cal L}}(z_m/\omega) 
		= \frac{1}{\omega^m} \sum_{i=1}^{m}C_i\cdot (2 \pi N)^i\ , 
\end{equation}
where the coefficients $C_i$ do not depend on $\omega$. Only the terms 
proportional to $N$ contribute to ${\cal L}_{\rm eff}$ and we obtain
\begin{equation} 
	{\cal L}_{\rm eff}^{(m-1)} = \frac {1}{\omega^{m-1}}\, C_1
\end{equation}
at order $m-1$.
Our choice of $\delta$ and the condition (\ref{FormAss}) assure
that we can truncate the series after a few terms.

To lowest order, we have
\[ {\cal L}_{\rm eff}^{(0)} = \frac {1}{N\tau}\int\limits_{0}^{N\tau} 
		\! dt_1\,\bar{{\cal L}}(t_1) \ .\]
This integral vanishes for $n \not= 1$. So, to lowest order 
the driving field only
contributes at resonance ($n = 1$), where we obtain the results of the
rotating wave approximation (see Section \ref{RWA}). 
A calculation of the first order corrections in the case of resonance 
($n=1$) yields
\begin{eqnarray}
	{\cal L}_{\rm eff} & = & {\cal L}_{\rm eff}^{(0)} + 
		{\cal L}_{\rm eff}^{(1)}\nonumber\\
	{\cal L}_{\rm eff}^{(1)} & = & \frac{\eta}{\omega} \left( 
		\begin{array}{cccc}
		0 & 0 & 0 & 0\\
		0 & 0 & -2 u v\eta & -\frac 1 2 u \delta \\
		\frac 1 2 \gamma u n^0 & 2 u v \eta & 0 &  \frac 3 4 u^2 
		\eta \\
		0 & \frac 1 2 u \delta & -\frac 3 4 u^2 \eta & 0 
	\end{array}\right)\ .
\end{eqnarray}
This can be expressed in terms of an effective Hamiltonian and 
an effective equilibrium state defined by the relation
\begin{equation}
	{\cal L}_{\rm eff} \bar{\rho} =: i [\bar{\rho},H_{\rm eff}] 
		- \gamma(\bar{\rho}-\rho^{eq}_{\rm eff})\ .
\end{equation}
We obtain
\begin{eqnarray}
	H_{\rm eff}& = & H_{\rm eff}^{(0)} + H_{\rm eff}^{(1)} \nonumber\\
	H_{\rm eff}^{(0)} & = & - \delta  \frac{\sigma_x}{2} + u 
		\eta \frac{\sigma_y}{2}\nonumber\\
	H_{\rm eff}^{(1)} & = & \frac{\eta}{\omega} \left[-\frac 3 4 
		u^2 \eta \frac{\sigma_x}{2} 
			-\frac 1 2 u \delta \frac{\sigma_y}{2} + 2 u v 
		\eta \frac{\sigma_z}{2}\right]
\end{eqnarray}
and
\begin{eqnarray} 
	\rho^{eq}_{\rm eff} & = &  \rho^{eq} + \rho_{\rm eff}^{(1)}
		\nonumber\\
	\rho_{\rm eff}^{(1)} & = & \frac{\eta}{\omega} \frac 1 2 u 
		n^0 \frac{\sigma_y}{2}\ .
\end{eqnarray}

We still have to calculate the evolution between
$N \tau$ and $t$. This can be done using perturbation theory. 
With ${\cal L}_1(t):=\bar{{\cal L}}(t)-{\cal L}_{\rm eff}$, we write  
the time evolution operator as the series 
\begin{eqnarray}\label{FormZeitentw}
	{\cal U}(t) & = & {\cal U}(t-N\tau){\cal U}(N\tau) \nonumber\\
	& = & e^{{\cal L}_{\rm eff} t} + \int\limits_{N\tau}^{t} \! dt_1\,
		e^{{\cal L}_{\rm eff} (t-t_1)} {\cal L}_1(t_1) 
		e^{{\cal L}_{\rm eff} t_1}\nonumber\\
	&&+ \int\limits_{N\tau}^{t} \! dt_1\!\!\int\limits_{N\tau}^{t_1} 
		\!dt_2 \, 
		e^{{\cal L}_{\rm eff} (t-t_1)}{\cal L}_1(t_1)
		e^{{\cal L}_{\rm eff} (t_1-t_2)} {\cal L}_1(t_2) 
		e^{{\cal L}_{\rm eff} t_2}
	+ \ldots\ .
\end{eqnarray}
Thus, we see that to lowest order the time evolution is completely
described by the effective Liouvillian, while the corrections are small 
due to our choice of $\delta$ and the condition (\ref{FormAss}).

In the first part of this paper, we presented a formalism in which we 
treated a harmonically driven TLS. We could map the time dependent 
problem onto a static one and give, order by order, an effective 
Liouvillian describing it. The lowest order result could be identified
with the RWA.

This formalism is not restricted to a simple TLS, but can be easily
generalised to any system driven by a periodic field. Thus we can
generalise the RWA to systems that have not a simple geometrical
interpretation like the TLS by identifying it with the lowest order
effective Liouvillian.
In the second part, we demonstrate this by investigating a coupled pair
of TLSs in a harmonic field.

\section{A Pair of Coupled Tunneling Systems}

Depending on the detailed nature of the TLSs in 
mixed crystals, an electric and an elastic moment 
is connected to the defect which can thus
interact with lattice vibrations, external fields or 
neighbouring defects. Consequently only at low defect
concentrations one can describe the situation by
isolated TLSs. With rising concentration 
pairs, triples etc. of defects are involved until
finally one faces a complicated many-body system
\cite{Aloisbuch,Orestispaper}.

Echo experiments \cite{sli} are an experimental tool to investigate
such systems \cite{paperrobert}.
In the following we investigate the influence of the interaction 
between the tunneling defects in such experiments.
Since we treat pairs of TLSs our results are confined to 
the limited range where the concentration is sufficiently high to
give a considerable amount of pairs but still below
the value where many-body effects set in.

We model a pair of coupled tunneling systems (TLSs) by the Hamiltonian
\begin{equation} 
	H_{\rm pair} = -\frac{k+\delta k}{2} \sigma_x\otimes 1 -
	\frac{k-\delta k}{2} 1\otimes\sigma_x -J \sigma_z\otimes\sigma_z. 
\end{equation}
The first and second term describe symmetric TLSs with $k>\delta k >0$,
while the third term models the dipole--dipole--interaction of the TLSs 
with the coupling constant $J$. 
The external harmonic force, coupling to dipole moment 
$p$ of the pair, is represented by 
\begin{equation} H_{\rm ext} = -\eta \sin(\omega t) (\sigma_z\otimes 1 
	+ 1\otimes\sigma_z ) \end{equation}
with the external frequency $\omega$ and the coupling constant $\eta$.
This ansatz assumes that the two TLSs are parallel to each other and to 
the harmonic force.
The full Hamiltonian of the problem reads 
$H=H_{\rm pair}+H_{\rm ext}$
and we are interested in the time-dependent dipole moment of the pair
assuming that we start from thermal equilibrium.
 
In the following, we outline the calculations, while
details can be found in Appendix \ref{Anhang_A}.
First, we diagonalize $H_{\rm pair}$ to find two {\sl uncoupled effective} 
TLSs:
\begin{equation} 
	\hat{H}_{\rm pair} = -\frac{\epsilon_1}{2} \sigma_x\otimes 1 -
	\frac{\epsilon_2}{2} 1\otimes\sigma_x, 
\end{equation}
with the energies $\epsilon_{1/2}=\sqrt{J^2+k^2}\,\pm\,
\sqrt{J^2+\delta k^2}$.
These effective TLSs have a more complicated coupling to the external field
than the original ones, reading
\begin{equation} 
	\hat{H}_{\rm ext} = -\eta \sin(\omega t) \left\{ (u_1 + u_2 \sigma_x)
		\otimes\sigma_z + \sigma_z\otimes (u_3 + u_4\sigma_x) \right \}
\end{equation}
where the $u$'s dependend on $k$, $\delta k$ and $J$ with values in the 
interval $[-1,1]$ (for further information see Appendix \ref{Anhang_diag}).

Now, we take the dominant effect of $\hat{H}_{\rm pair}$ 
into account by changing to a rotating frame of reference
%
\begin{equation}\label{rhodach} 
	\hat{\rho} (t) = \exp(i\omega t(n\,\sigma_x\otimes 1 + m\,1
		\otimes\sigma_x)/2) \; 
		\tilde{\rho} (t) \; \exp(-i\omega t(n\,\sigma_x\otimes 1 
		+ m\,1\otimes\sigma_x)/2)
\end{equation} 
where the integers $n$ and $m$ have to be chosen such that
\[ |\epsilon_1 - n\omega | < \frac{1}{2} \omega \quad \mbox{and} 
	\quad |\epsilon_2 - m\omega | < \frac{1}{2} \omega\ .
\]
Regarding only resonance phenomena we have to distinguish three cases.
The first two cases describe a situation where the internal frequencies
$\epsilon_1$ and $\epsilon_2$ are clearly separated. In the first case,
the external frequency $\omega$ is in resonance with 
the smaller internal frequency $\epsilon_2$ ($m=1$ and $n>1$), while
in the second case it is in resonance with $\epsilon_1$ ($m=0$ and $n=1$).
In the third case, $\omega$ is close to both internal frequencies ($m=n=1$).
This implies that both the energy separation $\delta k$ and the coupling $J$
of the underlying TLSs are small compared to their energy $k$, i.e.
$J,\delta k\ll k$.

We will discuss mainly the first case with $m=1$ and $n>1$. The results
for the second case are essentially the same and can be found in Appendix
\ref{Anhang_case2}. The treatment of the third case has also been relegated to
the appendix; in spite of more complicated calculations it gives no principally
new insight into the problem.

Neglecting dissipation ($\gamma = 0$), focusing on the situation of 
exact resonance ($\epsilon_2=\omega$) and following section \ref{Formalismus}, 
one obtains for the dipole moment to lowest order:
%
\begin{equation}\label{dipol} 
	p^{(0)}(t) = \frac{1}{4} \sum_{i=1,2} \frac{\Omega_i}{\eta} A_i \left\{ 
		\sin((\omega - \Omega_i) t) - \sin((\omega + \Omega_i) t) \right\}
\end{equation}
with
%
\begin{equation}\label{Rab} 
	\Omega_{1/2} = \eta(u_1\pm u_2) \qquad 
	A_{1/2} = r_2(1\pm r_1) \qquad 
	r_{1/2} = \tanh(\beta \epsilon_{1/2}/2) \ .
\end{equation}
Thus we find {\sl two} Rabi frequencies $\Omega_{1/2}$ yielding 
four frequencies for the dipole moment: $\omega_p=\epsilon_2\pm\Omega_{1}$ 
and $\omega_p=\epsilon_2\pm\Omega_{2}$. This is one of the main
results of our paper. Although the two TLSs can be transformed into
effective {\sl uncoupled} TLSs they do not give the same results as
truly uncoupled TLSs, which have only {\sl one} Rabi frequency.
%
%
\begin{figure}[t]
\centering{\epsfysize=7cm\epsffile{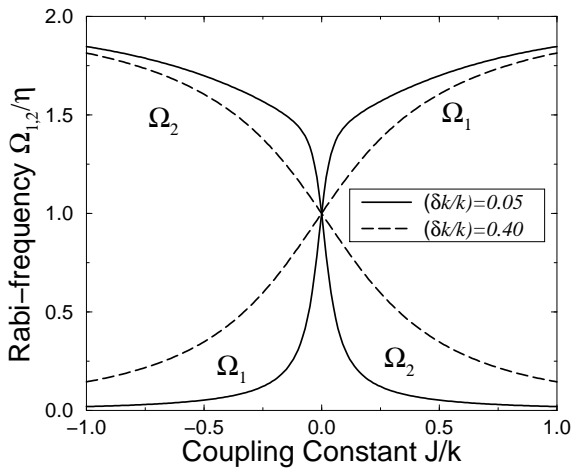}}
\caption{\label{Rab12}The two Rabi frequencies $\Omega_{1,2}$ for two
	different values of $(\delta k/k)$ against $(J/k)$}
\end{figure}
In Fig. \ref{Rab12} we have plotted the two Rabi frequencies 
(Eq. (\ref{Rab})) versus the coupling constant $J$.
At $J=0$, i.e. if the two spins are not coupled, the two Rabi frequencies 
coincide, thus reproducing the result of a single spin. 
For increasing $J$, $\Omega_1$ grows, tending quickly to $2 \eta$, whereas
$\Omega_2$ goes down to zero. In the case of negative $J$, $\Omega_1$ and
$\Omega_2$ interchange their roles. 

\begin{figure}[t]
\centering{\epsfysize=7cm\epsffile{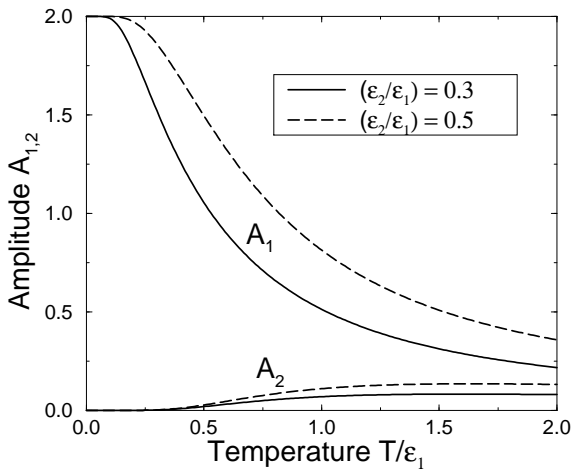}}
\caption{\label{Temp12}$A_{1/2}=r_2 (1 \pm r_1)$ for two
	different values of $(\epsilon_2/\epsilon_1)$ against $(T/\epsilon_1)$}
\end{figure}

The amplitudes of the peaks (Eq. (\ref{Rab})) are proportional to 
the frequencies themselves, thereby enhancing the signal of the 
bigger Rabi frequency. Additionally the amplitudes also contain 
a temperature dependent factor $A_{1/2}$ shown in Fig. \ref{Temp12}.
%
%
Over the whole temperature range, $A_1$ is larger than $A_2$. This
is most pronounced for very low temperatures where $A_1$ tends to
a value of 2 whereas $A_2$ vanishes. For higher temperatures starting
at about $T=\epsilon_1$ they are of the same order.

For $J>0$, $A_1$ corresponds to the higher Rabi frequency. In this case 
the higher Rabi frequency peak dominates the spectrum for all temperatures,
reaching its maximum for small temperature, while the
contribution of the lower Rabi frequency can be seen only at medium
temperatures. In the strong coupling limit the lower Rabi frequency
and thus also its amplitude vanish, leaving us with only one Rabi frequency 
of $2 \eta$.
This is in accordance with the picture that two strongly coupled 
TLSs form one small {\sl effective} TLS; its effective dipole moment
is the sum of the individual dipole moments, thus yielding 
$\Omega=2\eta$.

The latter situation has been investigated by
Weis et al. \cite{paperrobert}, 
who measured strongly coupled pairs of 
Lithium tunneling defects in a $KCl$ matrix. The tunneling defects, which can
be well described as TLSs \cite{Gomez,Peter}, interact via a 
dipole--dipole coupling. In the special case investigated by Weis et al. 
this coupling is ferromagnetic ($J>0$). 
Due to the strong coupling only one Rabi peak was found.
It would be desirable to investigate different pair constellations
corresponding to weaker or even antiferromagnetic couplings
in order to see the second Rabi frequency.

For $J<0$, $A_1$ corresponds to the lower Rabi frequency. Thus, the
amplitude of the higher Rabi frequency is suppressed by the temperature 
dependent factor and we can find a parameter regime, where both amplitudes
are of the same order of magnitude. For strong coupling and low
temperature, both amplitudes vanish, rendering corrections to the {\sl RWA}
of order $(\eta/\omega)$ important. These corrections lead to four
new peaks close to the bigger frequency $\epsilon_1$ (while we
are in resonance with $\epsilon_2$!) and can be found in Appendix 
\ref{Anhang_case1}. 
Yet due to the overall factor of $(\eta/\omega)$
we can not expect to see the new peaks in an experiment.

In the case of resonance with the larger frequency $\epsilon_1$, we
get essentially the same results, but the expressions for the two Rabi
frequencies and their amplitudes are different and are listed in 
Appendix \ref{Anhang_case2}. The main difference is the existence of
a parameter range
in which we get considerable amplitudes for both Rabi frequencies.
Furthermore, for $|J|\to\infty$ both Rabi frequencies vanish so that again
corrections to the RWA become dominant.

Concerning the third case where both inner frequencies are close 
to each other,
we only considered the situation of resonance with one of them while
$(\epsilon_1-\epsilon_2)\gg\eta$ (see Appendix \ref{Anhang_case2}). 
The results are very similar to the
previously discussed cases, but the corrections are of order 
$\eta/(\epsilon_1-\epsilon_2)$ rather than $(\eta/\omega)$.

\section{Conclusion}

In this paper we have communicated a generalisation of the
rotating wave approximation (RWA). First, we propose a formal
framework in which for a single spin the RWA appears as the
leading order. This formalism allows to easily calculate
higher order corrections to the approximation. If, on the other
hand, we apply it to a general $n$-level-system,
we can generalise the concept of the RWA by identifying it with 
the leading order contribution.

Second we have considered a coupled spin pair in an oscillating external 
field. Although the energetic structure of such a system 
is well described by two isolated effective spins the 
time evolution of the system shows some 'dynamical entanglement'. 
These effects will disappear in the ferromagnetic strong 
coupling limit, which makes clear that our results are 
consistent with previous investigations 
\cite{Aloisbuch,paperrobert,Orestispaper}. 

Furthermore our calculations also show that there is a 
situation (concerning spin pairs) where terms beyond the 
leading order of the RWA become important. This will be  
the case for an anti-ferromagnetic coupling, where the 
amplitudes of the leading order are suppressed. We
predict that whenever a situation can be modeled 
by a spin pair with such a coupling, rotary echos should 
show signals arising from higher order terms in the 
approximation scheme we proposed. 
This explicit example shows that, although the RWA 
seems to be correct for most cases, one can think of 
situations where the RWA does not suffice in order to 
separate the 'slow part' from the complete quantum dynamics.

{\bf ACKNOWLEDGEMENTS}
We wish to acknowledge helpful discussions with H. Horner, R. K\"uhn and 
H. Kinzelbach.
B. Thimmel is supported by a grant of the Graduiertenkolleg ``Physikalische 
Systeme mit vielen Freiheitsgraden'',
P. Nalbach is supported by the DFG-project HO 766/5-1 "Wechselwirkende 
Tunnelsysteme in Gl\"asern und Kristallen bei tiefen Temperaturen".
O. Terzidis thanks the French Foreign Office and the CROUS de Versailles 
for a postdoctoral grant. 
\pagebreak
\begin{appendix}
\section{Calculations and Corrections for the Pair}\label{Anhang_A}
\subsection{Diagonalization of the Pair Hamiltonian}
\label{Anhang_diag}
We diagonalize $H_{pair}$ by the transformation
\begin{equation}
	\hat{H}_{pair}=e^{-i(\frac{\alpha}{2} \sigma_y\otimes\sigma_z + 
		\frac{\beta}{2} \sigma_z\otimes\sigma_y)}
		\; H_{pair} \; e^{i(\frac{\alpha}{2} \sigma_y\otimes\sigma_z
		 + \frac{\beta}{2} \sigma_z\otimes\sigma_y)}\ ,
\end{equation}
where the angles $\alpha$ and $\beta$ obey
\begin{equation} 
	\tan(\alpha+\beta) = -\frac{J}{k} \quad \mbox{and}\quad
	\tan(\alpha-\beta) = -\frac{J}{\delta k} \ .
\end{equation}
This leads to the following values of $u$ in $H_{ext}$
\begin{equation}
	u_1 = \cos(\beta) \quad  , \quad u_2 = - \sin(\alpha) \quad , \quad 
	u_3 = \cos(\alpha) \quad , \quad u_4 = - \sin(\beta)\ .
\end{equation}
\subsection{First Case: $m=1$ and $n\ge 2$}
\label{Anhang_case1}
In the rotating frame of reference we calculate the effective Hamiltonian
following the procedure presented in section \ref{Formalismus}.
Including the dominant corrections of order $|\epsilon_1 - \omega|/\omega$,
the effective Hamiltonian reads 
%
\begin{equation} \label{Hklein1}
	\bar{H}_{\rm eff} =  -\frac{\delta_1}{2}\sigma_z\otimes 1 - 
		\frac{\delta_2}{2}1\otimes\sigma_z 
      + \frac{\eta}{2} (u_1 + u_2 \sigma_z)\otimes\sigma_y
	  + \frac{\delta_1}{\omega}\,\frac{\eta}{(n^2-1)}\sigma_y
		\otimes(u_3 + u_4 \sigma_z)\ ,
\end{equation}
where $\delta_1=\epsilon_1-n\omega$ and $\delta_2=\epsilon_2-m\omega$.
Using Eq. (\ref{FormZeitentw}), we can calculate the dipole moment up to
order $\eta/\omega$. Considering exact resonance, $\delta_2=0$, we
obtain
\begin{equation} 
	\begin{array}{rcl} p^{(1)}(t) & = &  \frac{1}{4} \{ 
		( \frac{\Omega_1}{\eta} A_1 + \frac{\eta}{\omega}B_1 )  
		\sin((\omega - \Omega_1) t) - 
		( \frac{\Omega_1}{\eta} A_1 - \frac{\eta}{\omega}B_1 ) 
		\sin((\omega + \Omega_1) t) \} \\
&&	+ \frac{1}{4} \{ 
		( \frac{\Omega_2}{\eta} A_2 + \frac{\eta}{\omega}B_1 )  
		\sin((\omega - \Omega_2) t) - 
		( \frac{\Omega_2}{\eta} A_2 - \frac{\eta}{\omega}B_1 ) 
		\sin((\omega + \Omega_2) t) \} \\
&& 	+ \frac{\eta}{\omega} B_2 \sin(\omega t) \\
&& 	- \frac{\eta}{\omega} B_3 \Big\{ \sin((\epsilon_1 + \eta u_2)t) 
		+ \sin((\epsilon_1 - \eta u_2)t) \Big\} \\
&& 	- \frac{\eta}{\omega} B_3 \Big\{ \sin((\epsilon_1 + \eta u_1)t) 
		+ \sin((\epsilon_1 - \eta u_1)t) \Big\} 
\end{array} \end{equation}
with
\[ \begin{array}{lll}
	B_1 = \left( 
	\frac{(u_1-u_2)^2}{4}-\frac{n}{n^2-1}u_3u_4\right) r_2(1-r_1) &
	B_2 = \frac{2 n}{n^2-1} (u_3^2+u_4^2) r_1 &
	B_3 = \frac{1}{n^2-1} u_3(u_3+r_2u_4)r_1\ ,	
	\end{array}
\]
while $A_{1/2}$ and $\Omega_{1/2}$ as well as $r_{1/2}$ were already 
defined in Eq. (\ref{Rab}).

\subsection{Second Case: $m=0$ and $n=1$}
\label{Anhang_case2}
\begin{figure}[t]
\parbox{7cm}{
\centering{\epsfysize=5.5cm\epsffile{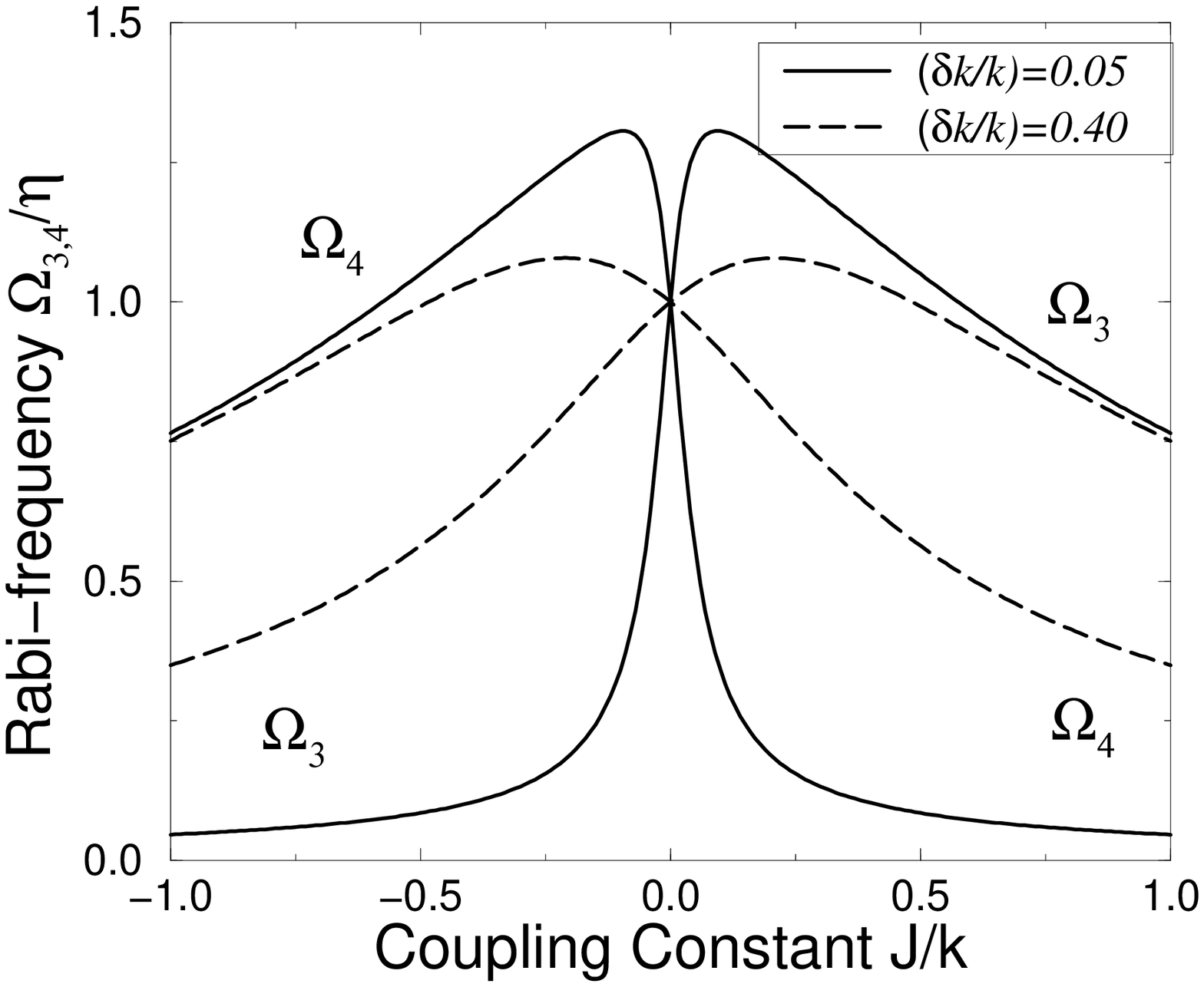}}
\caption{\label{Rab34}The two Rabi frequencies $\Omega_{3,4}$ for two
	different values of $(\delta k/k)$ against $(J/k)$}
}
\hfill
\parbox{7cm}{
\centering{\epsfysize=5.5cm\epsffile{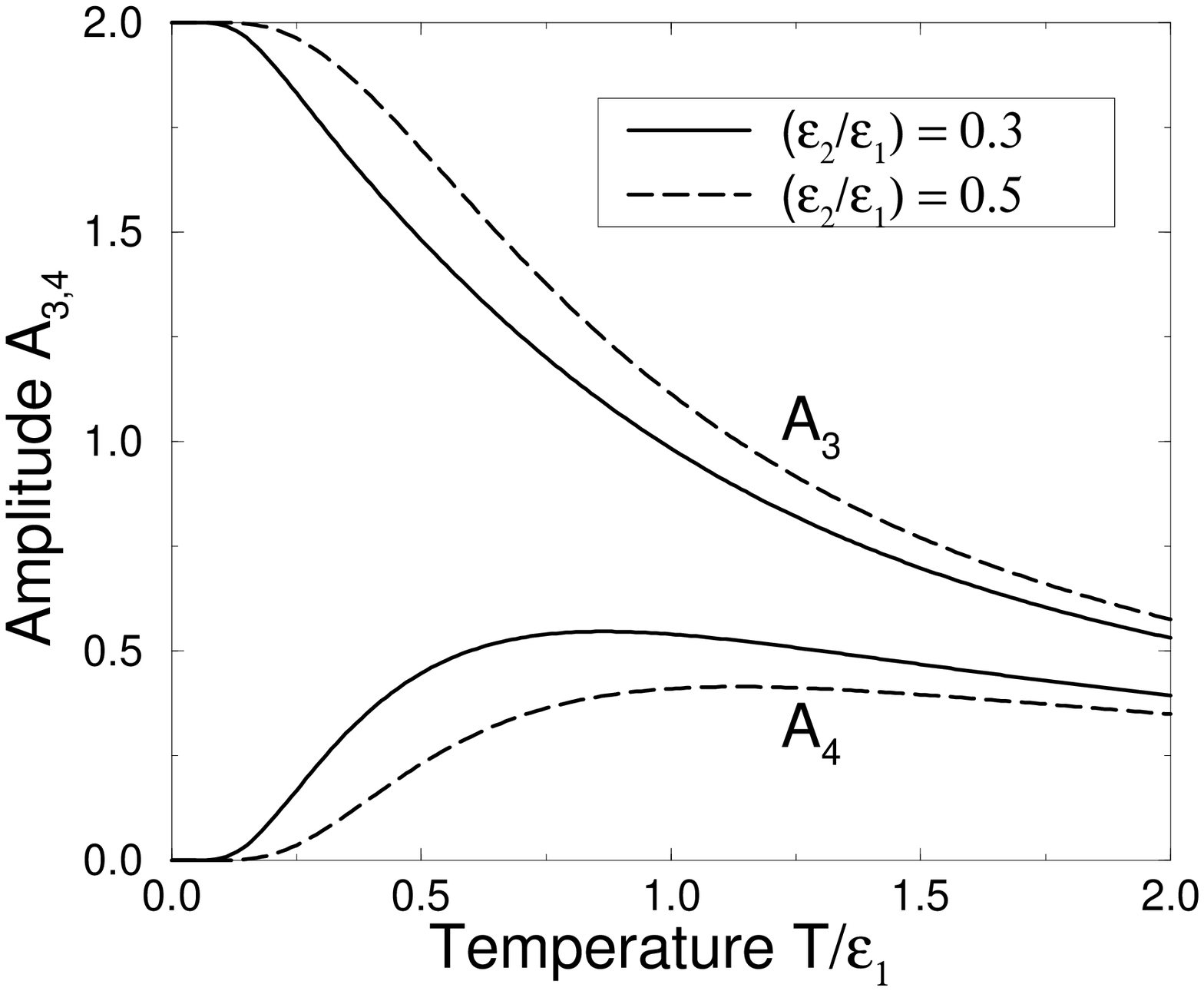}}
\caption{\label{Temp34}$r_1 (1 \pm r_2)$ for two
	different values of $(\epsilon_2/\epsilon_1)$ against $(T/\epsilon_1)$}
}
\end{figure}

All quantities can be obtained from those of the preceding case
using the following
transformations:
\begin{equation} 
	n\leftrightarrow m ,\quad r_1\leftrightarrow r_2 ,\quad 
	u_1\leftrightarrow u_3 \quad \mbox{and}\quad u_2\leftrightarrow u_4
\end{equation}
As the main results we show the Rabi frequencies plotted versus 
$J/k$ in Fig.\ref{Rab34} (corresponding to Fig.\ref{Rab12}) and 
the temperature dependent prefactors $A_i$ plotted versus 
temperature in Fig.\ref{Temp34} (corresponding to Fig.\ref{Temp12}).
%
%
\subsection{Third Case: $n=m=1$}
\label{Anhang_case3}
In this case, the effective Hamiltonian reads to lowest order
%
\begin{equation} \label{hnulldrei} 
	\bar{H}_{\rm eff} =  -\frac{\delta_1}{2}\sigma_z\otimes 1 -
		\frac{\delta_2}{2}1\otimes\sigma_z 
              + \frac{\eta}{2} \, \sigma_y\otimes(u_3 + u_4 \sigma_z)
			  + \frac{\eta}{2} \, (u_1 + u_2 \sigma_z)\otimes\sigma_y\ ,
\end{equation}
which is very similar to that in Eq.(\ref{Hklein1}). 
Restricting ourselves to the situation where the energy difference 
of the two effective TLS's $(\epsilon_1-\epsilon_2)$ is large compared to 
$\eta$, we can use perturbation theory.
If the bigger frequency is in resonance with the field, $\omega =\epsilon_1$,
we obtain for the dipole moment:
\begin{equation} 
	\begin{array}{rll} p^{(0)}(t) & = &  \frac{1}{4} \{ 
		( \frac{\Omega_3}{\eta} A_3 +2 \frac{\eta}{\delta_2}B_5 ) 
		 \sin((\omega - \Omega_3) t) - 
		( \frac{\Omega_3}{\eta} A_3 - 2\frac{\eta}{\delta_2}B_5 )
		 \sin((\omega + \Omega_3) t) \} \\
&&	+ \frac{1}{4} \{ 
		( \frac{\Omega_4}{\eta}A_4 - 2\frac{\eta}{\delta_2}B_6 ) 
		 \sin((\omega - \Omega_4) t) - 
		( \frac{\Omega_4}{\eta}A_4 + 2\frac{\eta}{\delta_2}B_6 ) 
		\sin((\omega + \Omega_2) t) \} \\
&& 	+ \frac{\eta}{\delta_2} B_7  \sin(\omega t)  \\
&& 	- \frac{\eta}{2\delta_2} B_8 \Big \{ \sin((\bar{\epsilon}_2 + \eta u_3)t) 
		+ \sin((\bar{\epsilon}_2 - \eta u_3)t)\Big \}\\ 
&& 	- \frac{\eta}{2\delta_2} B_9 \Big \{ \sin((\bar{\epsilon}_2 + \eta u_4)t) 
		+ \sin((\bar{\epsilon}_2 - \eta u_4) t)\Big \}
	\end{array} 
\end{equation}
with
\begin{equation}
	\begin{array}{lll} \Omega_{3/4} = \eta(u_3\pm u_4) & 
		A_{3/4} = r_1(1\pm r_2) & 
		B_{5/6} = r_1u_1u_2(1\pm r_2)\\
		B_7 = (u_1^2+u_2^2) r_2 & 
		B_{8/9} = r_2u_{2/1}(u_{1/2}r_1+u_{2/1}) & 
		\bar{\epsilon}_2 = \epsilon_2 + \frac{\eta^2}{2\delta_2}(u_1^2+u_2^2). 
	\end{array}
\end{equation}

\end{appendix}


\begin{thebibliography}{99}

\bibitem{Nara} V. Narayanamurti, R. O. Pohl, Rev. Mod. Phys. {\bf 42}, 
201 (1970)

\bibitem{Aloisbuch} A. W\"urger, From coherent tunneling to relaxation,
 Springer Tracts in Modern Physics 135, Springer Heidelberg (1996)

\bibitem{sli} C. P. Slichter, Principles of magnetic resonance, 
Springer series in solid state sciences 1, Springer (1996) 

\bibitem{RabOsc} I. I. Rabi, Phys. Rev. {\bf 51}, 652 (1937)

\bibitem{Blochgl} F. Bloch, Phys. Rev. {\bf 70}, 460 (1946)

\bibitem{Floquet} J. H. Shirley, Phys. Rev. {\bf 138}, B979 (1965)

\bibitem{Contfrac} S. H. Autler, C. H. Townes, Phys. Rev. {\bf 100},
 703 (1955)

\bibitem{BSshift} F. Bloch, A. Siegert, Phys. Rev. {\bf 57}, 522 (1940)

\bibitem{Gomez} M. Gomez, S. P. Bowen, J. A. Krumhansl, 
Phys. Rev. B {\bf 153}, 1009 (1967)

\bibitem{Peter} P. Nalbach, O. Terzidis, J. Phys. Condens. Matter 
{\bf 9}, p. 8561-77 ('97)

\bibitem{paperrobert} R. Weis, C. Enss, A. W\"urger, F. L\"uty,
 Ann.Phys {\bf 6}, p.263-86 ('97): Coherent Tunneling of Lithium
 Defect Pairs in KCl Crystals 
\newline 
R. Weis, Doktorarbeit, Ruprecht-Karls-Universit\"at Heidelberg 1995

\bibitem{Orestispaper} O. Terzidis, A. W\"urger, Journal of 
Physics Cond. Matt. {\bf 8},  7303 (1996) 
\newline  
O. Terzidis, Doktorarbeit, Ruprecht-Karls-Universit\"at Heidelberg 1995

\end{thebibliography}
\end{document}